\documentclass[10pt]{article}
\usepackage[margin=1in]{geometry}
\usepackage{amsmath}
\usepackage{abstract}
\usepackage{amsfonts,color}
\usepackage{amssymb}
\usepackage{graphicx}
\usepackage{hyperref}
\usepackage{bigints}
\usepackage{mathrsfs}
\usepackage[toc]{appendix}
\begin{document}
\title{\bf{Lagrangian Description of Accreting Black Hole Systems in the Context of Emergent Spacetime}}
\author
{Satadal Datta $^\ast$, Tapas K. Das $^\dagger$\\
{\small\textit{Harish-Chandra Research Institute, HBNI, Chhatnag Road, Jhunsi, Allahabad-211019, India}}\\
%$^2$Homi Bhabha National Institute, Training School Complex,\\
%Anushaktinagar, Mumbai - 400094, India\\
\date{}
\small Email: $^\ast$ satadaldatta1@gmail.com, satadaldatta@hri.res.in,\\ \small$^\dagger$tapas@hri.res.in}
\maketitle
\begin{abstract}
We make use of the Lagrangian description of fluid motion to highlight certain features in the context of spacetime geometry as emergent phenomena in fluid systems. By using Lagrangian Perturbation Theory (LPT), we find that if the flow is not barotropic it will not be locally irrotational, and as a result emergent gravity can not be realised for such flow. Our work gives a new perspective of examining the perturbations in a fluid from a different approach (other than Eulerian approach) which is the way of using Lagrangian Perturbation Theory.  We make use of Lagrangian description of motion to examine the propagation of Eikonal wave (wave having very short wavelength) from the reference frame of the observer moving with the background flow. We find that waves of ultra short wavelength propagate similar to waves in a static uniform medium in the near vicinity of the observer. We restrict ourselves to nonrelativistic flows in astrophysical black hole accretion.
\end{abstract}
\section{Introduction}
Linear perturbations on a steady state background irrotational inviscid flow give rise to a field equation satisfied by the linear perturbation in velocity potential which has striking similarities with a massless scalar field equation in a curved spacetime \cite{a}\cite{b}\cite{c}. If the background steady flow is transonic, analogy with a blackl hole or a white hole spacetime can be drawn \cite{b}. If the flow becomes supersonic from subsonic, analogy with a black hole horizon comes into light, and in the other case, analogy with a white whole spacetime is visible. In that context, linear perturbation is introduced on the density field and the velocity field in the flow; the whole approach is done by treating density and velocity as fields which is the essence of Eulerian description of fluid motion. Here we explore the Lagrangian description of motion to describe the phenomena from a different point of view and we use LPT \cite{d}, \cite{e} to find certain restrictions on the perturbation itself to mimic massless KG field equation in a curved spacetime. As we know that for fluid, compressibility property is an important feature. Due to the compressibility property of an ideal fluid, thermodynamics of a fluid is determined from the dynamics of the flow itself. Thus in a fluid in motion, dynamical variables (velocity, position etc) and therdonynamical variables (pressure, density etc) are connected due to this property. This property makes the system interesting to study. Barotropic (the condition of pressure being only function of density) flow is a subset of compressible flow indeed. This is a necessary requirement in physical acoustics for studying emergent gravity. We use LPT and we find that if the flow is not barotropic, it will not be locally free from rotationality, and thus making it impossible to study emergent gravity in such a flow in the regime of physical acoustics. Our work gives a novel perspective of looking into the problem from a different angle, i.e., the the use fo LPT in such a system to study. In this work, we consider large scale realistic systems, i.e, black hole accretion. We work with the models of accreting black hole, and we introduce a procedure to find a rough estimate about the wavelength of Eikonal waves \cite{w}, \cite{x} originated at different positions, such that moving (with the background flow) observers at those positions in the medium, observe the wave to propagate just like waves in a uniform static medium (in the neighbourhood of those observers).
\section{Lagrangian Description of Fluid Motion}
In the Lagrangian description of fluid motion \cite{f}, instead of using fluid density and velocity as fields, one follows the motion of a fluid element \cite{f}. Inviscid fluid equations in Lagrangian description are given by \cite{f}
\begin{equation}
\frac{d\rho}{dt}+\rho{\bf{\nabla}}.{\bf{v}}=0
\end{equation}
\begin{equation}
\frac{d{\bf{v}}}{dt}=-\frac{{\bf{\nabla}}p}{\rho}-{\bf{\nabla}}\Psi({\bf{x}})
\end{equation} 
where $\rho,~{\bf{v}},~p,~\Psi$ are fluid density, velocity, fluid pressure and scalar potential corresponding to external body force respectively. $\frac{d}{dt}$ is Lagrangian time derivative \footnote{Lagrangian time derivative is also denoted by $\frac{D}{Dt}$ in some text books and the literature.}. The first equation describes the conservation of mass in a fluid element in motion and the second one describes the equation of motion by Newton's law of motion.\\
The position coordinate of fluid element is given by ${\bf{x}}({\bf{R}},t)$ where ${\bf{x}}({\bf{R}},0)={\bf{R}}$. The velocity of the element is
\begin{equation}
\dot{{\bf{x}}}({\bf{R}},t)={\bf{v}}
\end{equation}
'Dot' means $\frac{d}{dt}$. 
Using equation (2) and another initial condition on velocity, one can uniquely find the position of a fluid element in the flow as a function of time.\\
Let consider a steady flow, i.e. $\frac{\partial (~) }{\partial t}=0$. In the steady state flow, we denote the  pressure field, density field and velocity field by $p_0, \rho_0$ respectively. The velocity vector of a fluid element by ${\bf{V}}({\bf{R}},t)$  , satisfying
\begin{equation}
\frac{d{\bf{V}}}{dt}=-\frac{{\bf{\nabla}}p_0}{\rho_0}-{\bf{\nabla}}\Psi({\bf{x}})
\end{equation}
Position of the fluid element is ${\bf{X}}(t)$ and $\dot{{\bf{X}}}(t)={\bf{V}}$. 
\section{Linear Perturbations}
We introduce linear perturbations in the fluid quantities as follows,
\begin{align*}
& p({\bf{x}}, t)=p_0({\bf{x}})+p'({\bf{x}}, t)\\
&\rho({\bf{x}}, t)=\rho_0({\bf{x}})+\rho'({\bf{x}}, t)\\
&{\bf{v}}({\bf{x}}, t)={\bf{v}}_0({\bf{x}})+{\bf{v}}'({\bf{x}}, t)\\
\end{align*}
The Eulerian perturbations are denoted by $p',~\rho'$ and ${\bf{v}}'$. $|p'|<<p_0,~|\rho'|<<\rho_0$ and magnitude of ${\bf{v}}'$ is also small. \\
The Lagrangian perturbations in LPT are related to the Eulerian perturbations via another vector field, called Lagrangian displacement, ${\bf{\delta}}({\bf{x}}, t)$. ${\bf{\delta}}$ represents the displacement of fluid elements in space from their position of equilibrium, ${\bf{X}}(t)$s. The Lagrangian perturbations in the first order of smallness are given by \cite{d}, \cite{f}
\begin{align*}
& \Delta p=p'+\delta.{\bf{\nabla}}p_0\\
& \Delta \rho=\rho'+\delta.{\bf{\nabla}}\rho_0\\
& \Delta{\bf{v}}=\frac{d{\bf{\delta}}}{dt}=\frac{\partial\delta}{\partial t}(={\bf{v}}'({\bf{x}}, t))+\delta.{\bf{\nabla}}{\bf{v}}_0
\end{align*}
where $\Delta p$ and $\Delta \rho$ are related by
\begin{equation}
\frac{\Delta p}{p_0}=\gamma\frac{\Delta \rho}{\rho_0}~{\rm or}~\frac{\Delta p}{\Delta\rho}=c_{s0}^2
\end{equation} 
where $c_{s0}$ is the thermodynamic sound speed in the medium, $\gamma$ is the specific heat ratios, $\gamma=1$ if the perturbation is isothermal in nature. For air, sound propagates adiabatically \cite{f}, i.e., no heat transfer occurs between adjacent volume elements.\\   
We write inviscid irrotational fluid equations in Eulerian description as
\begin{equation}
\frac{\partial\rho}{\partial t}+{\bf{\nabla}.(\rho{\bf{v}})}=0
\end{equation}
\begin{equation}
\frac{\partial \bf{v}}{\partial t}+{\bf{v}}.{\bf{\nabla}}{\bf{v}}=-\frac{{\bf{\nabla}}p}{\rho}-{\bf{\nabla}}\Psi({\bf{x}})
\end{equation}
\begin{equation}
{\bf{\nabla}}\times{\bf{v}}=0
\end{equation}
Defining velocity potential as ${\bf{v}}=-{\bf{\nabla}}\psi$, we find the Euler equation for the perturbed quantity as
\begin{equation}
-{\bf{\nabla}}\frac{\partial \psi'}{\partial t}+{\bf{\nabla}}({\bf{v}}_0.{\bf{v'}})=-\frac{{\bf{\nabla}}p'}{\rho_0}+\frac{\rho'}{\rho_0^2}{\bf{\nabla}}p_0
\end{equation}
Now from equation (5) and from the expression of Lagrangian perturbations,
\begin{equation}
p'=c_{s0}^2\rho'+\left(c_{s0}^2-\frac{dp_0}{d\rho_0}\right)\delta.{\bf{\nabla}}\rho_0
\end{equation}
Now if the background medium has different kind of stratification than the nature of perturbation, the term in the right hand side of equation (9), would involve an extra quantity ${\bf{\delta}}$ and the term in the right hand side can not be written as gradient of a quantity, i.e, enthalpy in the perturbed medium can not be defined and as a result the motion would not be irrotational in that case, evident from equation (9). For example, let us consider a medium of isothermal stratification and the propagating disturbance to be adiabatic in nature, therefore the sound speed is $c_{s0}=\sqrt{\frac{\gamma p_0}{\rho_0}}$ and $\frac{dp_0}{d\rho_0}=\frac{p_0}{\rho_0}=\frac{1}{\gamma}c_{s0}^2$, the second term in the right hand side, in equation (10) does not vanish. Similarly, for isothermal sound propagating in a medium of adiabatic stratification, the same thing happens. \\
If the back ground medium has same kind of stratification as the nature of disturbance, from equation (10),
\begin{equation}
p'=c_{s0}^2\rho'
\end{equation}
From equation (9),
\begin{equation}
\partial_t\psi'={\bf{v}}_0.{\bf{v'}}-\frac{p'}{\rho_0}
\end{equation}
Now after some manipulations one can find the field equation for $\psi'({\bf{x}},t)$; given by
\begin{equation}
\partial_\mu(f^{\mu\nu}({\bf{x}})\partial_\nu)\psi'({\bf{x}},t)=0
\end{equation}
where 
\begin{equation}
f^{\mu\nu}({\bf{x}})\equiv\frac{\rho_0}{c_{s0}^2}\begin{bmatrix}
-1 & \vdots & -v'^{j} \\
\cdots&\cdots&\cdots\cdots \\
-v'^{j}&\vdots & c_{s0}^{2}\delta^{ij}-v'^{i}v'^{j}
\end{bmatrix}
\end{equation}
Now comparing with massless scalar field equation in a curved spacetime in general, one can find the analogue acoustic metric by using $f^{\mu\nu}=\sqrt{-g}g^{\mu\nu}$. Then one can write down the acoustic metric as
\begin{equation}
ds^2=g_{\mu\nu}dx^{\mu}dx^{\nu}=\frac{\rho_0}{c_{s0}}\left(-(c_{s0}^2-v_{0}^2)dt^2-2{\bf{v}}_0dt.d{\bf{x}}+d{\bf{x}}^2\right)
\end{equation}
Therefore, the emergent spacetime feature through the perturbations in a steady flow can be realized if and only if the perturbation's nature matches with the stratification of the background medium \footnote{There is another possibility, if the medium is uniform, in that case the emergent spacetime metric is flat; here we are considering the medium to be stratified in general.}. Hence the emergent phenomena is restricted to isothermal perturbation in a isothermal background medium or adiabatic perturbation in a adiabatic background medium. Therefore, the flow has to remain barotropic in nature even in the presence of perturbation for the acoustic analogue of spacetime geometry to emerge.\\
Linear perturbation of Bernoulli's constant, $\zeta\left(=\frac{1}{2}{\bf{v}}^2+\int\frac{dp}{\rho}+\Psi({\bf{x}}\right)$ is $\zeta'$, related to $\psi'$ by $\partial_t \psi'=\zeta'$, obvious from equation (7) for barotropic flow. Therefore, linear perturbation of Bernoulli's constant, defined for barotropic flow \cite{f}, also gives rise to emrgent spacetime phenomena \cite{g}, \cite{h}. For which, the above conclusion is also true when analogue spaacetime is realized through the linear perturbation of Bernoulli's constant.\\
As we have assumed the the flow to be irrotational, Bernoulli's constant is a conserved quantity for the steady flow $({\bf{\nabla}}\zeta_0=0)$, therefore, $\Delta \zeta=\zeta'$.
We have
\begin{align*}
\partial_\mu(f^{\mu\nu}({\bf{x}})\partial_\nu)\zeta'({\bf{x}},t)=0
\end{align*}
\begin{equation}
\Rightarrow \partial_\mu(f^{\mu\nu}({\bf{x}})\partial_\nu)\Delta\zeta ({\bf{x}},t)=0
\end{equation}
Hence, the phenomena of emergent spacetime can also be realized through the Lagrangian perturbation of Bernoulli's constant. Lagrangian perturbation of Bernoulli's constant gives the change in the energy content (as Bernoulli's constant, being the sum of specific kinetic energy, potential energy and specific enthalpy, has the dimension of energy) carried by per unit mass of a fluid element in flow. Equation (16) describes the variation of this energy content per unit mass of fluid elements at different positions with time. As the fluid elements, undergoing through compression-expansion, oscillate back and forth with displacement ${\bf{\delta}}$ from the moving equilibrium position, ${\bf{X}}(t)$, the energy is transferred from one fluid element to the neighbours with the local speed of sound, discussed in details in the next section. If the flow is transonic in nature, black hole or white hole spacetime, depending on the direction of flow, corresponds to emergent spacetime \cite{i}. Equation (16) implies that for analogue blackl hole spacetime, as a fluid element crosses the sonic horizon \cite{j} from subsonic to supersonic region, the extra energy variation per unit mass of that element, due to linear perturbation, can never come back to the subsonic region. If a finite amount of energy is expended to perturb the flow by creating disturbances in the medium, the energy is distributed and transferred through the different elements in the medium and as the different fluid elements cross the analogue blackl hole sonic horizon from subsonic region to supersonic region and thus the fluid elements get lost through the sink (similar to the black hole singularity); in time, a part of the energy, the part carried and transferred by the elements along downstream, will be engulfed by the acoustic analogue of black hole. Only the energy which propagates upstream, the part of the total which is not totally carried by the fluid elements rather is being transferred from one element to the neighbours in the opposite direction of flow in the subsonic region of flow, would be available.  In the supersonic region of flow, the speed of the medium itself surpasses the local speed of sound, the speed at which energy is transferred from one fluid element to the neighbours at a given location, therefore no energy variation can escape sonic horizon. Similar but opposite conclusion is true for analogue white hole spacetime.
\section{Coordinate Transformation}
Let's follow the equilibrium position of an element, in the other words, the location of a fluid element in the absence of any disturbance. The equilibrium position vector is denoted by ${\bf{X}}(t)$ in general. Let the equilibrium position of a particular fluid element be denoted ${\bf{X}}({\bf{R}},t)$ where ${\bf{X}}({\bf{R}},0)= {\bf{R}}$. This notation indeed uniquely specify a particular fluid which was at ${\bf{R}}$ at $t=0$; and at a given time two fluid elements can not be in the same position. The velocity is ${\bf{V}}({\bf{R}},t)=\dot{{\bf{X}}}({\bf{R}},t)$. So far, we have described the motion in a coordinate system $({\bf{x}},t)$ which is rest in absolute space \cite{k} or moving with uniform velocity with respect to the absolute space or which is stationary with the source or sink (if exists) of the system; so that Newton's law is valid in the reference frame. Now we try to describe things from the equilibrium position of a particular fluid element which is accelerating in general due to external body force and pressure imbalance in the system. Coordinate of any point in the system with respect to the new coordinate system is $({\bf{x}}',t')$. $({\bf{x}}',t')$ is related to $({\bf{x}},t)$ via Galilean transformation \cite{l}, given by
\begin{equation}
{\bf{x}}'= {\bf{x}}-{\bf{X}}({\bf{R}},t)={\bf{x}}-\int^t {\bf{V}}({\bf{R}},t) dt-{\bf{R}}
\end{equation}
\begin{equation}
t'=t
\end{equation}
\begin{equation}
\frac{d{{\bf{x}}}'}{dt'}=\frac{d{{\bf{x}}}}{dt}-{\bf{V}}({\bf{R}},t)
\end{equation}
Therefore, using chain rule of partial derivatives, one can find
\begin{equation}
\frac{\partial}{\partial t}=\frac{\partial}{\partial t'}-{\bf{V}}({\bf{R}},t').{\bf{\nabla}}'
\end{equation}
\begin{equation}
{\bf{\nabla}}={\bf{\nabla}}'
\end{equation}
As a result, fluid equations, relating the density field and the velocity field, in this new coordinate system \footnote{The transformation is passive here, it does not change the field rather it changes the coordinate to describe those fields.} can be written in Eulerian description as
\begin{equation}
\frac{\partial\rho}{\partial t'}-{\bf{V}}({\bf{R}},t').{\bf{\nabla}}'\rho+\bf{\nabla}'.(\rho{\bf{v}})=0
\end{equation}
\begin{equation}
\frac{\partial \bf{v}}{\partial t'}+\left({\bf{v}}-{\bf{V}}({\bf{R}},t')\right).{\bf{\nabla}}'{\bf{v}}=-\frac{{\bf{\nabla}}'p}{\rho}-{\bf{\nabla}'}\Psi({\bf{x'}})
\end{equation}
In our case, the flow is irrotational, therefore
\begin{equation}
{\bf{\nabla}}'\times {\bf{v}}=0
\end{equation}
For the steady flow, $\frac{\partial ()}{\partial t}=0$, therefore we have 
\begin{equation}
\frac{\partial\rho_0}{\partial t'}-{\bf{V}}({\bf{R}},t').{\bf{\nabla}}'\rho_0=0
\end{equation} 
\begin{equation}
\Rightarrow \bf{\nabla}'.(\rho_0{\bf{v}}_0)=0
\end{equation}
\begin{equation}
\frac{\partial \bf{v}_0}{\partial t'}-{\bf{V}}({\bf{R}},t').{\bf{\nabla}}'{\bf{v}_0}=0
\end{equation}
\begin{equation}
\Rightarrow {\bf{v}_0}.{\bf{\nabla}}'{\bf{v}_0}=-\frac{{\bf{\nabla}}'p_0}{\rho_0}-{\bf{\nabla}'}\Psi({\bf{x'}})
\end{equation}
Therefore, for steady flow the fluid equations (equation(26) and equation (27)) in $({\bf{x}}',t')$ and $({\bf{x}},t)$ are same in form. \\
Linear Eulerian perturbations over the steady flow are introduced in the fluid system (section 3) in the $({\bf{x}},t)$ coordinate system, the equation for Eulerian perturbation fields in the $({\bf{x}}',t')$ coordinate system, are given by
\begin{equation}
\frac{\partial \rho'}{\partial t'}-{\bf{V}}({\bf{R}},t').{\bf{\nabla}}'\rho'+\bf{\nabla}'.(\rho'{\bf{v}}_0+\rho_0{\bf{v}}')=0
\end{equation}
\begin{equation}
\frac{\partial {\bf{v}}'}{\partial t'}+({\bf{v}_0}-{\bf{V}}).{\bf{\nabla}}'{\bf{v}}'+{\bf{v}}'.{\bf{\nabla}}'{\bf{v}_0}=-\frac{{\bf{\nabla}}'p'}{\rho_0}+\frac{\rho'}{\rho_0^2}{\bf{\nabla}}'p_0
\end{equation}
One can easily see that again by doing coordinate transformation to $({\bf{x}},t)$, one recovers the original equations of the perturbations.\\
If the nature of perturbation and the stratification of the background medium are of same kind, one can find the emergent spacetime metric in this new coordinate, as follows
\begin{equation}
ds^2=g_{\mu\nu}({\bf{x}}',t')dx'^{\mu}dx'^{\nu}=\frac{\rho_0}{c_{s0}}\left(-\left(c_{s0}^2-({\bf{v}}_{0}-{\bf{V}})^2\right)dt'^2-2({\bf{v}}_0-{\bf{V}})dt'.d{\bf{x}}'+d{\bf{x}}'^2\right)
\end{equation}
The metric is no more time independent because $({\bf{x}}',t')$ frame is moving with respect to $({\bf{x}},t)$ frame.  The same metric can be directly derived from equation (15) and using the above coordinate transformation.\\
Sound wave of very short wavelength, i.e, in the eikonal limit, follows null geodesic and is insensitive to the conformal factor (see Appendix \ref{A}), described by the acoustic metric in the geometric limit, given by
\begin{equation}\label{Ometric}
ds^2|_{{\rm geometric}}={\tilde{g}_{\mu\nu}}({\bf{x}}',t')~dx'^{\mu}dx'^{\nu}=\left(-\left(c_{s0}^2-({\bf{v}}_{0}-{\bf{V}})^2\right)dt'^2-2({\bf{v}}_0-{\bf{V}})dt'.d{\bf{x}}'+d{\bf{x}}'^2\right)
\end{equation}
with the null geodesic condition, given by
\begin{equation}
ds^2|_{{\rm geometric}}=0.
\end{equation}
As the wavelength of the sound wave is very small, an observer moving with the fluid element can examine the wave within a very short radius around him such that ${\bf{v}}_{0}\approx {\bf{V}}$. Therefore, the acoustic metric perceived by them in the near vicinity around them will be
\begin{equation}
ds^2|_{{\rm geometric}}=-c_{s0}^2dt'^2+d{\bf{x}}'^2. 
\end{equation}
As the frequency of such wave is very high, within a short period of time, the observer moving with the element, will perceive $c_{s0}$ to be independent of time. Therefore, in the eikonal limit, an observer moving with a fluid element (more precisely moving with the background velocity of the medium), perceives sound in the neighbourhood of him/her as sound moving in a uniform medium. Within, small distance from that observer, the wavefront of the sound wave is not only plane but also does not change orientation, i.e., within short distance of the observer, the sound propagates in a particular direction. This is exactly similar to the principal of equivalence in General Theory of Relativity \cite{v} that a freely falling observer feels no gravity. Therefore, under the above coordinate transformation, in the near vicinity of the fluid element, the emergent
spacetime metric corresponds to acoustic analogue of Minkowski spacetime. Now as the fluid element moves in $(x, t)$ spacetime, the above coordinate transformation describes the
coordinate transformation to the local inertial frames \cite{v} at $X(R, t)$ at different time. Similarly, in the
very near vicinities of different fluid elements in motion, the emergent spacetime is flat. Therefore, this reference frame is similar to the local inertial frame in general Theory of relativity.
\section{Estimation of the Wavelength of the Perturbation}
We have discussed in the Appendix \ref{A} that eikonal wave does not see the conformal factor in the acoustic metric, therefore, we have in the frame of the observer moving with the background speed of the medium; the metric,
\begin{equation}
ds^2|_{{\rm geometric}}={\tilde{g}_{\mu\nu}}({\bf{x}}',t')~dx'^{\mu}dx'^{\nu}=\left(-\left(c_{s0}^2-({\bf{v}}_{0}-{\bf{V}})^2\right)dt'^2-2({\bf{v}}_0-{\bf{V}})dt'.d{\bf{x}}'+d{\bf{x}}'^2\right)
\end{equation}
In the near vicinity of the observer $\bf{v}_0\sim \bf{V}$, therefore, we can set a very small quantity; 
\begin{equation}
|\bf{v}_0-\bf{V}|=\epsilon\mathscr{V},
\end{equation}
where $\epsilon$ is a small dimensionless number $(<<1)$ and $\mathscr{V}$ is a quantity having dimension of speed, we choose this quantity according to the speed scale of the problem. We estimate the measure of the length around the observer over which the background speed of the medium differ by $\epsilon\mathscr{V}$, this length $l$ gives the measure of the wavelength of the eikonal wave. Therefore,  in the near vicinity of the observer, the acoustic metric of the emergent spacetime is effectively flat for waves having such short wavelength ($\leq l$). In other word, the speed of sound is same in all direction within this length $l$, and to realize such effect the wavelength of the linear disturbance has to be $\leq l$.
\section{Conical Adiabatic Flow}
The conical disk model is preferred than other models. Therefore, we choose to work with this model under post Newtonian potentials representing Schwarzschild \cite{e1} (we call it $\Phi_{\rm Schwarzschild}$) and Kerr spacetime \cite{f1}, \cite{g1} (we call it $\Phi_{\rm Kerr}$). We scale radial distance by Schwarzschild radius $(r_g=\frac{2GM_{BH}}{c^2})$ and potential (energy) by $c^2$, where $M_{BH}$ is the mass of the accretor and $c$ is the light speed in vacuum. 
There is a good discussion about the conical disk flow in the literature \cite{y}, \cite{g1}. Here, we have mentioned some additional details for the completeness of the work.
We scale speed by the light speed ($=c$), density by $\frac{M_{BH}}{r_g^3}$, pressure by $\frac{M_{BH}c^2}{r_g^3}$ and angular momentum by $cr_g$. Therefore, we work with dimensionless variables. 
In a conical flow, accreting fluid falls under gravity, having velocity components along radial and azimuthal direction, with axial symmetry. The rotating fluid falls into the star/accretor under the gravitational pull of the accretor through a channel having a solid $\Theta$. The gravity of the medium is not taken into account, i.e, we are using test fluid approximation in this chapter. The continuity equation can be written as
\begin{equation}
\frac{\partial \rho}{\partial t}+\frac{1}{rH}\frac{\partial }{\partial r}(rH\rho v)=0
\end{equation}
where $\rho$ and $v$ are the fluid density and radial velocity having spherical symmetry.
$H$, the height of the disk, is proportional to $r$. Therefore, for a steady flow,we have the expression of the conserved quantity along the flow, i.e, the mass accretion rate
\begin{equation}
{\dot{M}}=\Theta\rho v r^2,
\end{equation}
derived from the steady state continuity equation, given by
\begin{equation}
\frac{d(\rho v r H)}{dr}=0.
\end{equation}
Euler momentum equation is given by
\begin{equation}
\frac{\partial v}{\partial t}+v\frac{\partial v}{\partial r}=-\Phi '(r)-\frac{1}{\rho}\frac{\partial p}{\partial r}+\frac{\lambda^2}{r^3}
\end{equation} 
where $\Phi'(r)$ is the external field term and in this case it is gravitational force per unit mass of fluid exerted  by the accretor. $\lambda$ is the specific angular momentum of the fluid having small value such that viscosity is negligible \cite{d1}.
Therefore for a steady flow, we have, another conserved quantity, i.e, the Bernoulli's constant, given by
\begin{equation}
\zeta=\frac{1}{2}v^2+\int\frac{dp}{\rho}+\Phi(r)+\frac{\lambda^2}{2r^2},
\end{equation}
derived from the equation of motion, given by
\begin{equation}
v\frac{dv}{dr}=-\frac{1}{\rho}\frac{dp}{dr}-\Phi'(r)+\frac{\lambda^2}{r^3}.
\end{equation}
We consider adiabatic relation between pressure and density as
\begin{equation}
p=K\rho^\gamma.
\end{equation}
$K$ is a constant, a function of specific entropy \cite{x}.
Therefore, using the  barotropic equation and the above steady state equations, we have
\begin{eqnarray} \label{dvdr}
&\frac{dv}{dr}=\frac{\frac{2c_s^2}{r}+\frac{\lambda^2}{r^3}-\Phi'(r)}{v-\frac{c_s^2}{v}},\\
&\frac{dc_s}{dr}=c_s(1-\gamma)\left(\frac{1}{2v}\frac{dv}{dr}+\frac{1}{r}\right). \label{dcsdr}
\end{eqnarray}
We seek transonic solution. Therefore, there exists a finite radius at which $v_0=c_{s0}$. From, the expression of $\frac{dv}{dr}$, for the physical existence of such a finite radius (=$r_c$), the numerator of the equation \ref{dvdr} has to be zero because already the denominator is zero at that radius.
Sound speed at the critical radius
\begin{equation}\label{sc}
c_{sc}=\sqrt{\frac{r_c\Phi'(r_c)}{2}-\frac{\lambda^2}{2r_c^2}}
\end{equation}
Therefore, we have from the expression of $\zeta$ and from the aforementioned condition,
\begin{equation}
\zeta-\frac{\lambda^2}{2r_c^2}-\Phi(r_c)-\frac{\gamma+1}{4(\gamma-1)}\left[r_c\Phi'(r_c)-\frac{\lambda^2}{r_c^2}\right]=0.
\end{equation}
Therefore, for a given value of $\zeta$, $\gamma$, $\lambda$ under a gravitational potential, we can find the critical radius (the `dumb hole' horizon or the sonic horizon) \cite{d1}, \cite{y}. In general
\begin{equation}\label{rc}
r_c=r_c(\zeta,\gamma,\lambda).
\end{equation}
From the expression \ref{sc}, for a physical transonic solution to exist,
\begin{equation}
\Phi'(r_c)>\frac{\lambda^2}{r_c^3}.
\end{equation}
To find the steady state solution for the transonic accretion, we first find $\frac{dv}{dr}$ at the critical radius, $r_c$. Using L'Hospital's rule for finding value of a function at a $x$ value where it has $\frac{0}{0}$ form, we get
\begin{equation}
\frac{dv}{dr}|_{r=r_c}=q,
\end{equation}
with $q$ satisfying a quadratic equation, as follows
\begin{equation}
q^2+Bq+C=0.
\end{equation} 
The coefficients are given by
\begin{eqnarray}
& B=\frac{4c_{sc}(\gamma-1)}{(\gamma+1)r_c},\\
& C=\frac{1}{\gamma +1}\left[\frac{2c_{sc}^2}{r_c^2}+\frac{3\lambda^2}{r_c^4}+\Phi''(r_c)+\frac{4c_{sc}^2(\gamma-1)}{r_c^2}\right].
\end{eqnarray}
Using equation \ref{dcsdr}, one can find the value of $\frac{dc_s}{dr}$ at the critical point. Therefore, using these values at the critical point as initial conditions, one can find the steady state solution in general in the conical model (for more details with the other models, see \cite{y}, \cite{g1}).\\
Now we take the gravitational potential, $\Phi=\Phi_{\rm Schwarzschild}$ and $\Phi_{\rm Kerr}$, i.e., we consider two cases.
For the potential $\Phi_{\rm Schwarzschild}$, we find critical point, from the equation (the equation \ref{rc}),
\begin{eqnarray}\label{rca0}
& 8\zeta(\gamma-1)r_c^2(r_c-1)^2-(\gamma +1)r_c^3+2\lambda^2(\gamma+1)(r_c-1)^2+4(\gamma-1)r_c^2(r_c-1)\nonumber\\
&-4\lambda^2(\gamma-1)(r_c-1)^2=0.
\end{eqnarray}
For the potential $\Phi_{\rm Kerr}$, we get
\begin{eqnarray}\label{rca}
& 4\zeta(\gamma-1)\left[2\zeta r_1(1-\beta)+1\right]r_c^{3-\beta}(r_c-r_1)^\beta \nonumber\\
& -r_c^2\left[(\gamma+1)r_1(1-\beta)+4(\gamma-1)(r_c-r_1)\right]\nonumber\\
&+2\lambda^2 r_1(1-\beta)(3-\gamma) r_c^{1-\beta}(r_c-r_1)^\beta =0.
\end{eqnarray}
With $a=0$ equation \ref{rca} reduces to equation \ref{rca0}. Thus, we can find the steady state transonic solution under the considered gravitational potentials.\\
In reference to the work \cite{g}, we can write the acoustic metric in geometric limit, as
\begin{equation}\label{ds2gm}
ds^2|_{{\rm geometric}}=\left(-(c_{s0}^2-v_{0}^2)dt^2+2v_0(r)dtdr+dr^2\right),
\end{equation} 
where $r$ is cylindrical polar coordinate. Therefore, we can design the acoustic metric in such flow depending on the parameter set: $[\zeta,~\gamma,~\lambda,~a]$.
Let us consider an observer, moving along the $x-$ axis towards the accretor sitting at the origin. The observer is moving with the background speed of the medium. Therefore,according to the Galilean transformation:
\begin{equation}
x= x'+R-\int^t V(R,t) dt,
\end{equation}
\begin{equation}
t=t',
\end{equation}
\begin{equation}
y=y',
\end{equation}

As the observer moves towards the accretor, $V(R,t)$ takes the value of the speed of the background  medium. The initial distance of the observer is $R$. 
After making a coordinate transformation of the acoustic metric to the Cartesian coordinate in two dimension (because the radial flow is happening on the disk plane), we have ($c_{s0}$ and $v_0$ are scalar by definition)
\begin{eqnarray}
& ds^2|_{{\rm geometric}}=-(c_{s0}^2-v_0^2)dt^2+2v_0dt(cos\phi dx+sin\phi dy)\nonumber\\
& +(cos^2\phi dx^2+sin^2\phi dy^2+sin2\phi dxdy),
\end{eqnarray}
where 
\begin{eqnarray}
& cos\phi=\frac{x}{\sqrt{x^2+y^2}}\\
& sin\phi=\frac{y}{\sqrt{x^2+y^2}}.
\end{eqnarray}
As the observer is moving along the $x-$ axis, in the near vicinity of the observer, $\phi$ is roughly zero, and also as we are considering wave which propagates along the radial direction. The observer examines wave propagating on the $x-$ axis, therefore
\begin{equation}
ds^2|_{{\rm geometric}}=-(c_{s0}^2-v_0^2)dt^2+2v_0dtdx+dx^2,
\end{equation}
From the observer's reference frame, the above acoustic metric becomes
\begin{equation}
ds^2|_{{\rm geometric}}=-(c_{s0}^2-(v_0-V)^2)dt'^2+2(v_0-V)dt'dx'+dx'^2
\end{equation}
In the near vicinity of the observer, along the $x-$ axis $|v_0-V|=\epsilon \mathscr{V}$. Therefore, according to the discussion in the previous section,
\begin{equation}
ds^2|_{{\rm geometric}}=-(c_{s0}^2-\epsilon^2\mathscr{V}^2)dt'^2+2\epsilon\mathscr{V}dt'dx'+dx'^2.
\end{equation}
Hence, considering the leading terms in the acoustic metric in the near neighbourhood of the observer,
\begin{equation}
ds^2|_{{\rm geometric}}=-c_{s0}^2 dt'^2+dx'^2.
\end{equation} Eikonal wave follows null geodesic, hence it propagates with same speed $c_{s0}$ along positive $x$ axis and negative $x$ axis in the close vicinity of the observer. Therefore, in the near vicinity of the observer, the emergent spacetime is flat. Owing to the axial symmetry of the problem, any observer moving radially towards the accretor, will have same conclusion while studying radially propagating wave. Now we estimate the lengthscale $l$ over which the speed of the background medium varies by $\epsilon\mathscr{V}$ to get an idea about the wavelength of Eikonal wave at different radii to produce such effect. Therefore, it all depends on how $v_0(r)$ varies with $r$. Since ${\bf \nabla}={\bf \nabla}'$, we only try to find the variation in the reference frame fixed with respect to the accretor at the origin. The faster the variation of $v_0$ with $r$, smaller the length, $l$ is. In the next section we numerically compute $l$ by considering suitable $\epsilon\mathscr{V}$.
\section{Numerical Estimation of the wavelength}
As we choose $\epsilon\mathscr{V}$ small enough, therefore we can relate the length $l$ with step size (step length to numerically solve the problem) of the problem $h$ by the following manner;
\begin{equation}\label{lc}
l=\frac{h\epsilon\mathscr{V}}{\Delta v},
\end{equation}
where $\Delta v$ is the magnitude of change in background speed for the change of radial distance by $h$. We work with magnitudes of change to get an idea about $l$. $l=lc$ at the critical point (critical point and sonic point are same in the case of conical flow. We compute the variation of $log(l/lc)$ with radial distance. We present the variation in the following figures. We denote the critical point by C in red in the following figures. We also show the speed variation of the background medium with the radial distance to get an idea about how speed variation influences $l$. In velocity profile, if there is/are a turning point/s, we choose the minimum value of variation (taking into account variation of speed from both the increasing side and decreasing side of $r$ around the turning point/s) in background speed around the turning point to set $\epsilon\mathscr{V}$. If there is no turning point in the background velocity profile, we simply set $\epsilon\mathscr{V}=0.001\times v_c$, where $v_c$ the speed of the background medium at the sonic point. It is quite evident that one would get a good idea about the variation of $l$ with $r$ by choosing this methodology. We denote the turning points in the figure as $T1,~T2..$ We then vary the parameters of the problem $\zeta$ and $\lambda$ to see the variation in $l$ at different regions of the parameter space \cite{y}. First we work with pseudo-Schwarzschild potential and then we consider Pseudo-Kerr potential. 
 
\begin{figure}[hbtp]
\centering
%\captionsetup{margin=0.8cm}
\includegraphics[scale=0.38]{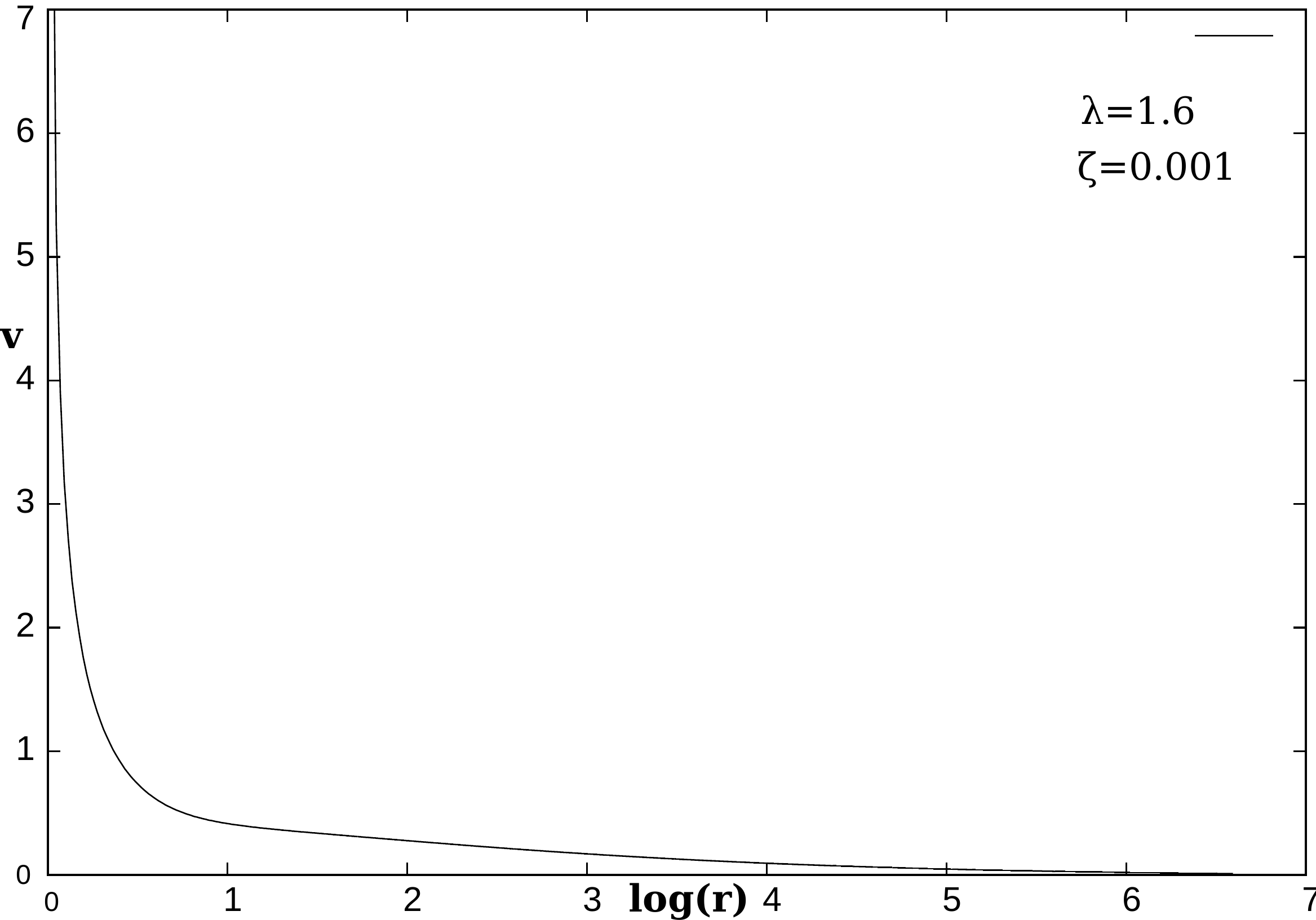}\\\includegraphics[scale=0.38]{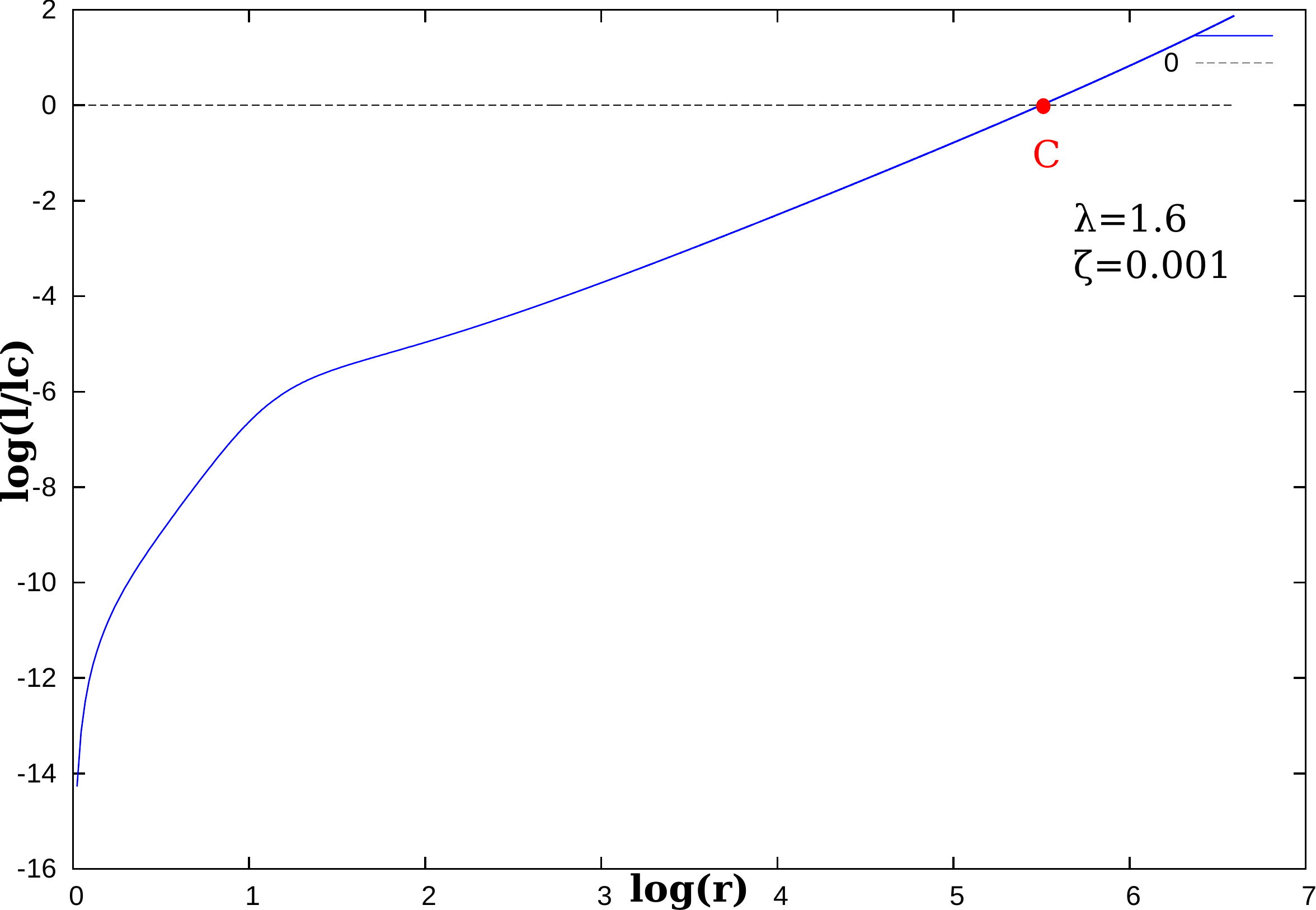}  
%\label{Figure1}
\caption{Background speed variation (top) and variation in $l$ (bottom). Region in the parameter space: region with single critical point relatively away from the accretor.}
%justification=centering,
%\includegraphics[width=0.325\textwidth]{satadal_f1.pdf}
\end{figure}
\begin{figure}[hbtp]
\centering
%\captionsetup{margin=0.8cm}
\includegraphics[scale=0.38]{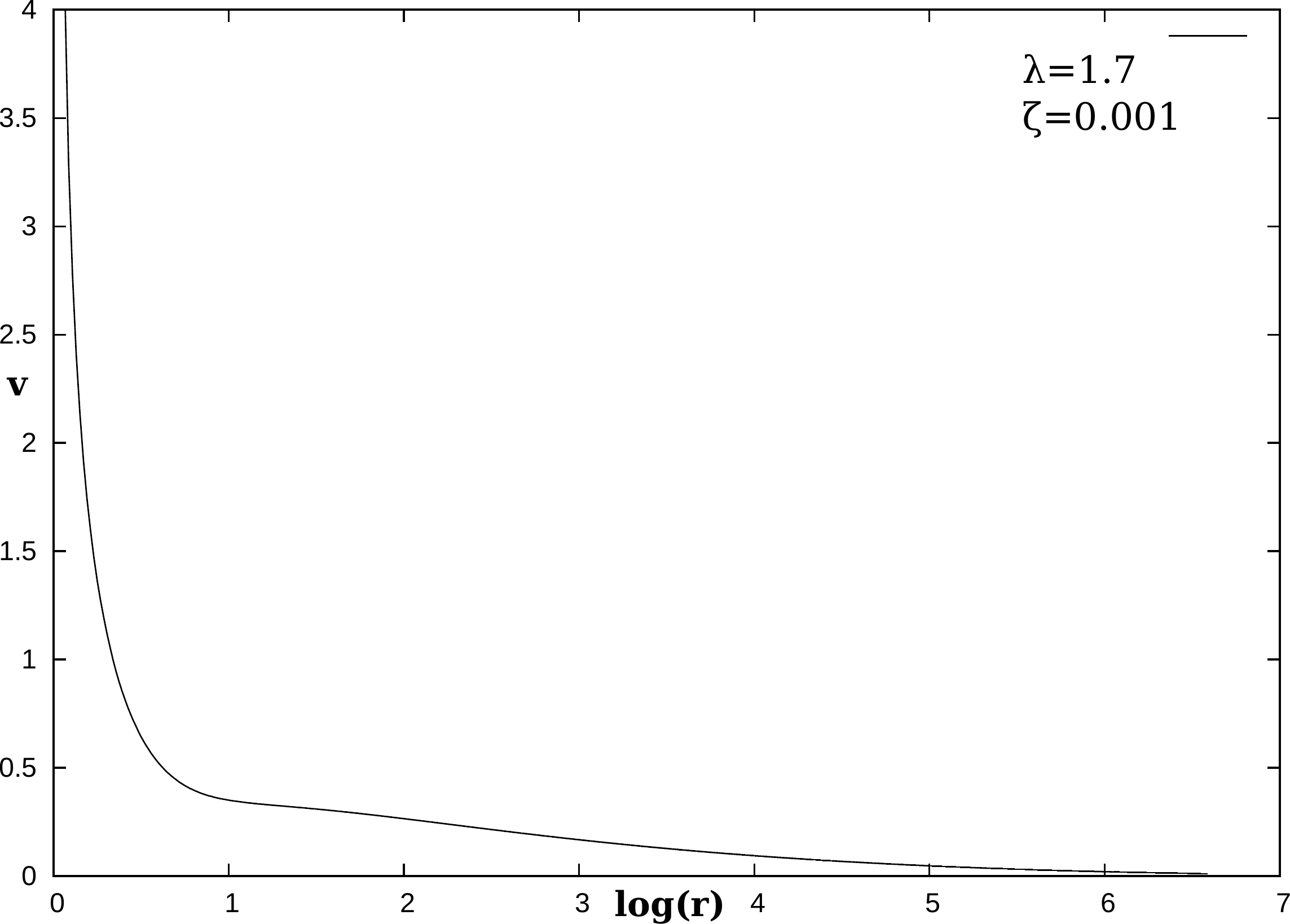}\\\includegraphics[scale=0.38]{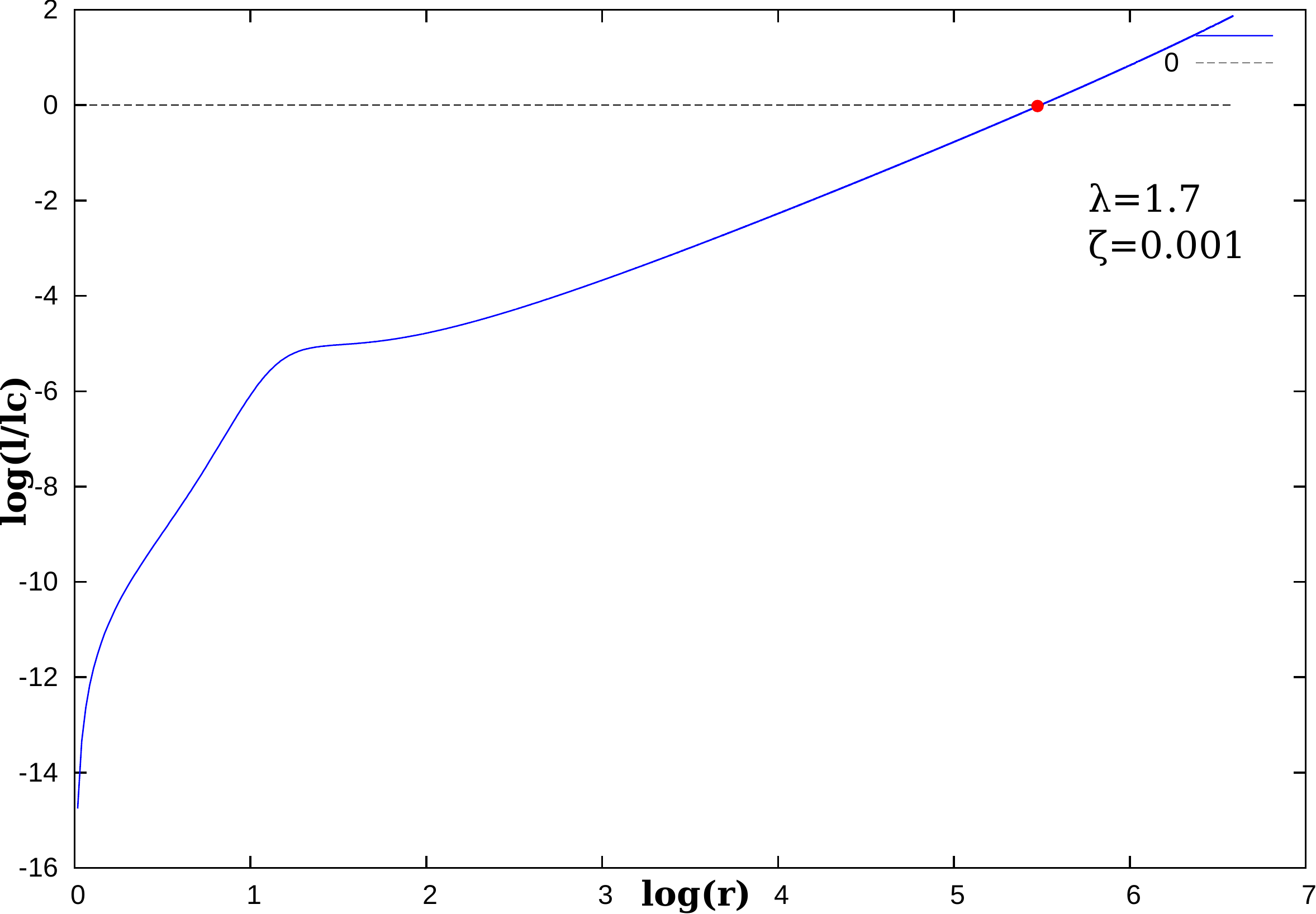}  
%\label{Figure1}
\caption{Background speed variation (top) and variation in $l$ (bottom). Region in the parameter space: region with single critical point situated relatively away from the accretor.}
%justification=centering,
%\includegraphics[width=0.325\textwidth]{satadal_f1.pdf}
\end{figure}
\begin{figure}[hbtp]
\centering
%\captionsetup{margin=0.8cm}
\includegraphics[scale=0.38]{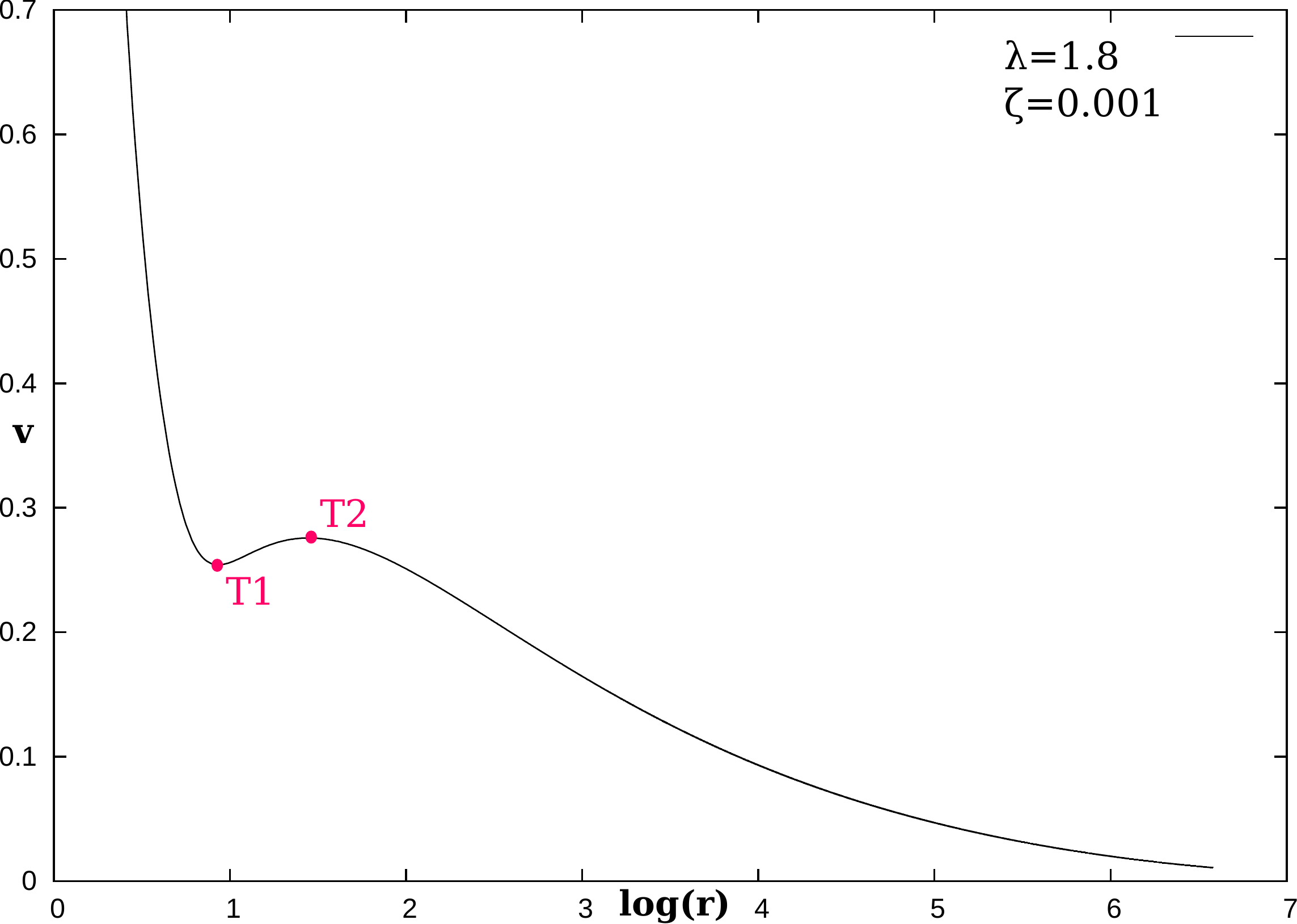}\\\includegraphics[scale=0.38]{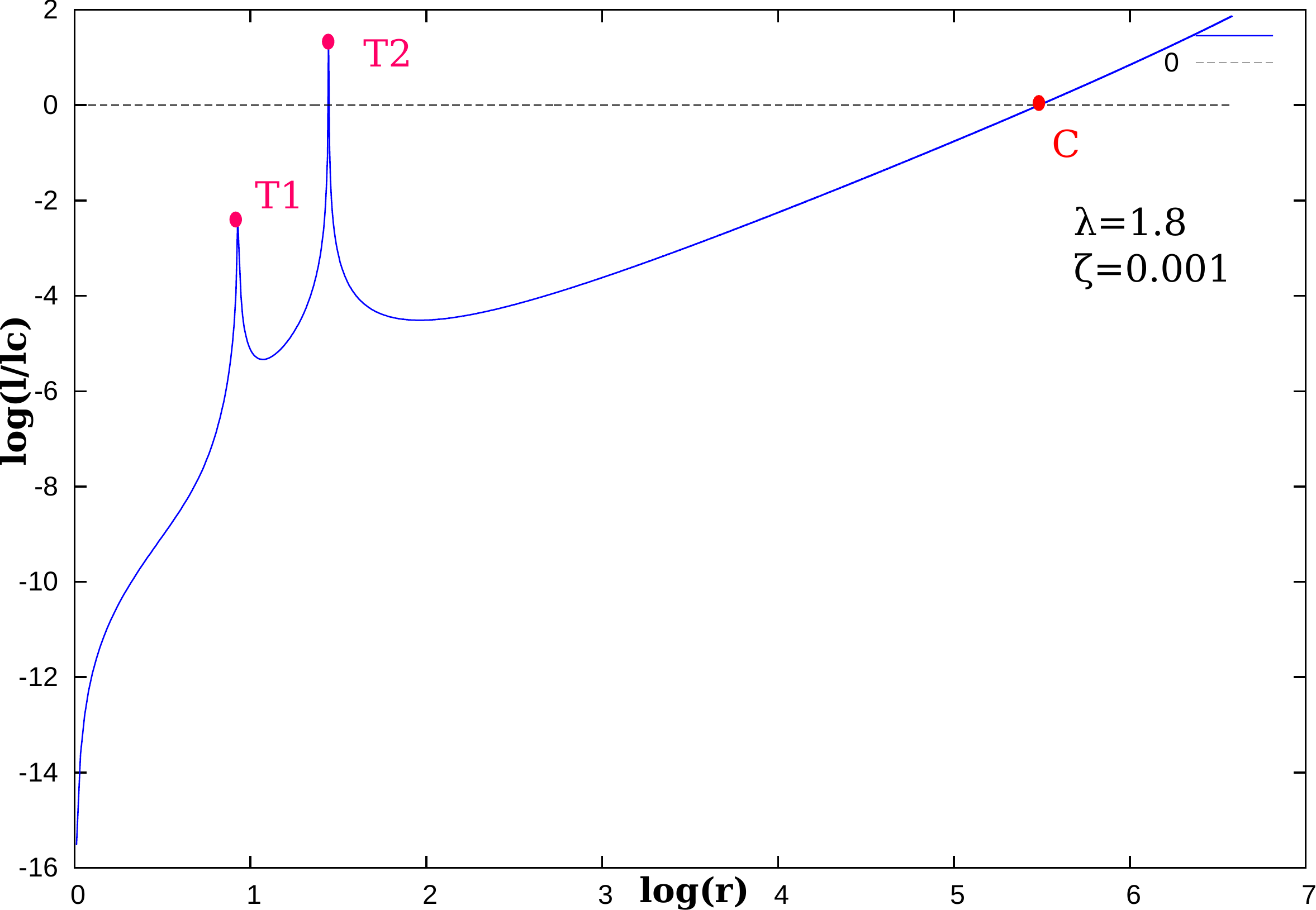}  
%\label{Figure1}
\caption{Background speed variation (top) and variation in $l$ (bottom). Region in the parameter space: region with three critical points, transonic accretion curve passing through the farthest (from the accretor) critical point. Formation of sharp peaks at the turning points in the figure of $l-r$ variation.}
%justification=centering,
%\includegraphics[width=0.325\textwidth]{satadal_f1.pdf}
\end{figure}
\begin{figure}[hbtp]
\centering
%\captionsetup{margin=0.8cm}
\includegraphics[scale=0.38]{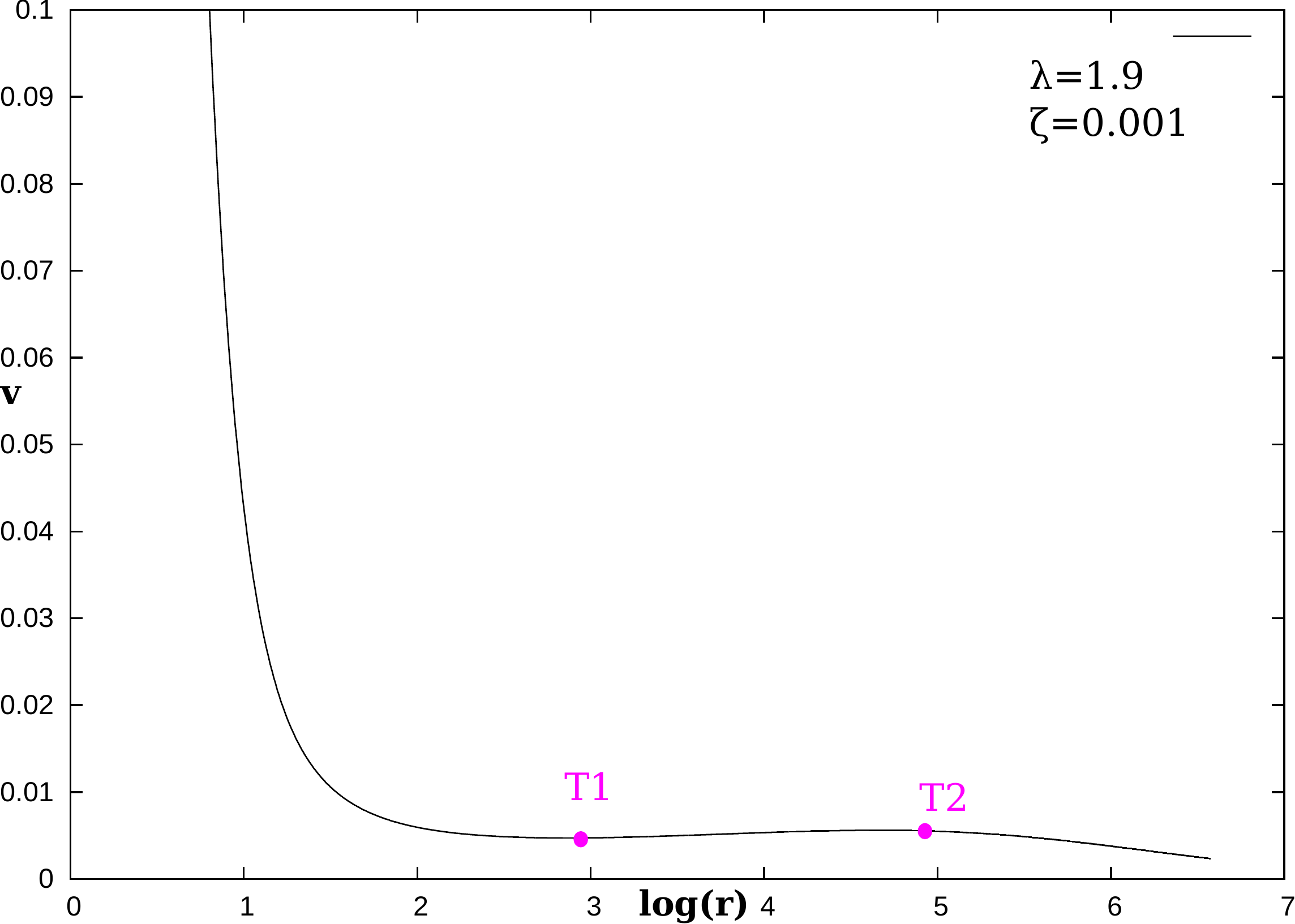}\\\includegraphics[scale=0.38]{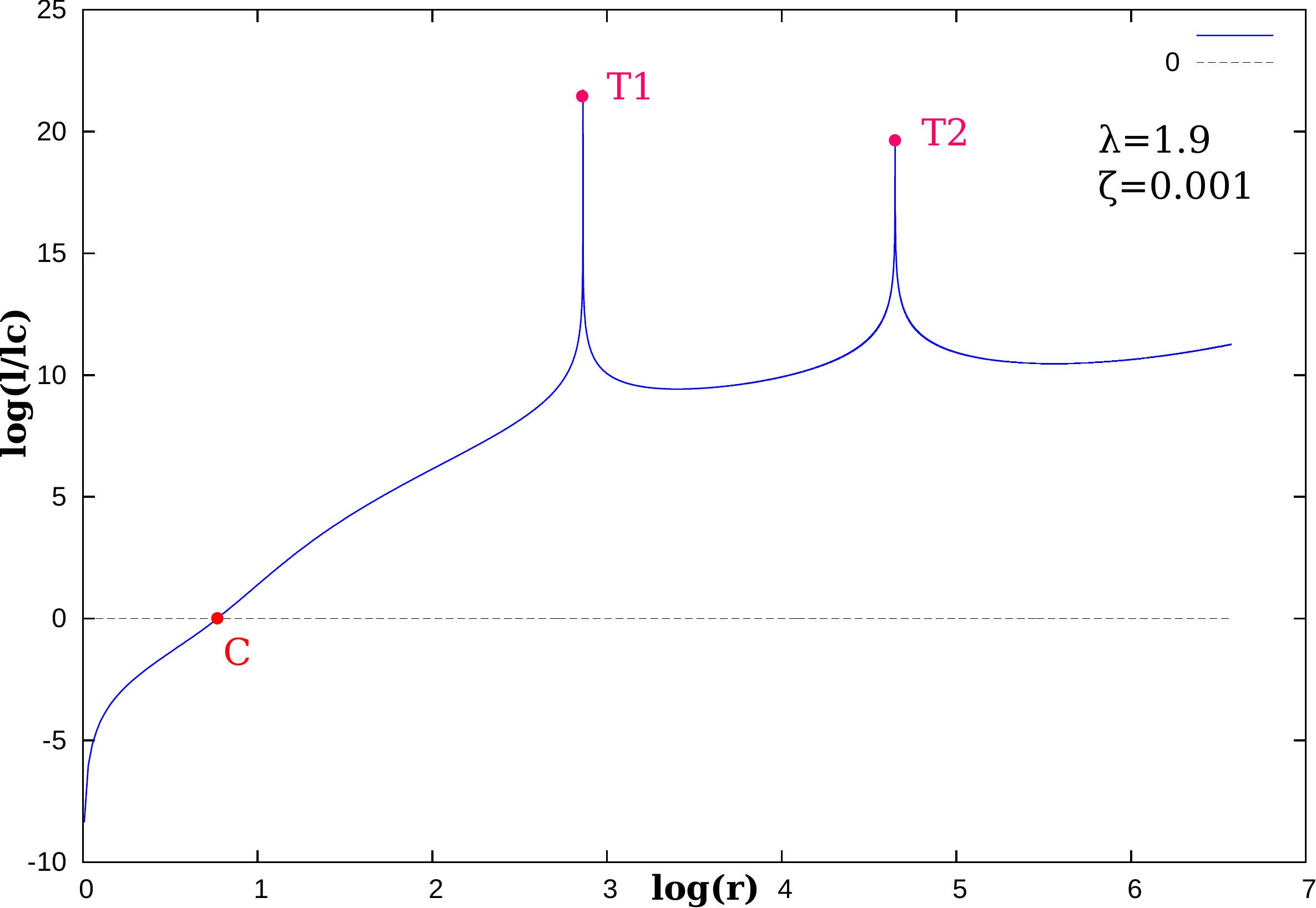}  
%\label{Figure1}
\caption{Background speed variation (top) and variation in $l$ (bottom). Region in the parameter space: region with three critical points, transonic accretion curve passing through the nearest (from the accretor) critical point. Formation of sharp peaks at the turning points in the figure of $l-r$ variation.}
%justification=centering,
%\includegraphics[width=0.325\textwidth]{satadal_f1.pdf}
\end{figure}
\begin{figure}[hbtp]
\centering
%\captionsetup{margin=0.8cm}
\includegraphics[scale=0.38]{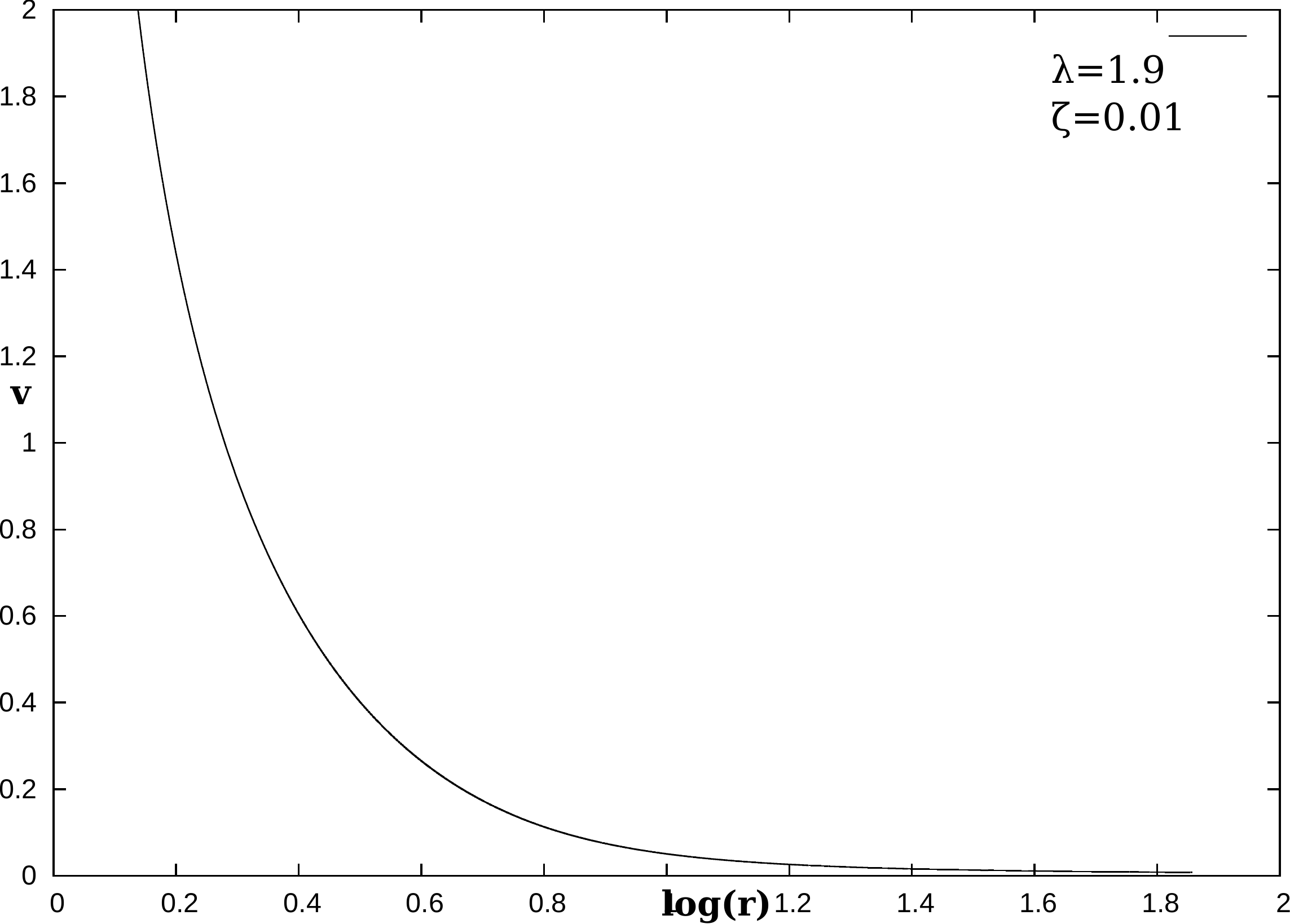}\\\includegraphics[scale=0.38]{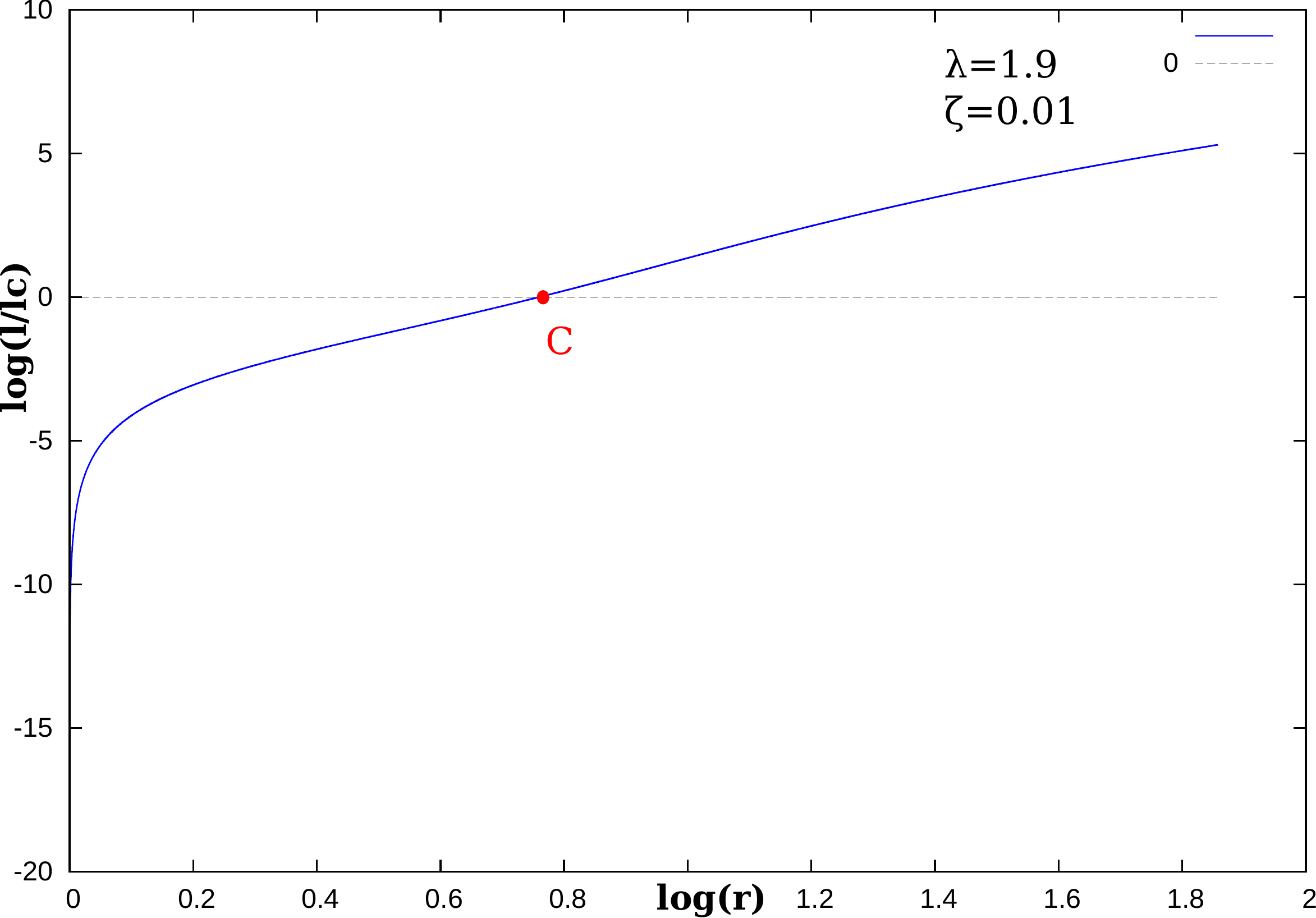}  
%\label{Figure1}
\caption{Background speed variation (top) and variation in $l$ (bottom). Region in the parameter space: region with single critical point situated near the accretor.}
%justification=centering,
%\includegraphics[width=0.325\textwidth]{satadal_f1.pdf}
\end{figure}
Thus we cover each single region of the parameter space. The peaks form at the turning points because the variation of the background speed becomes increasingly slow around the turning points, thus resulting into sharp increase in the length $l$. We can categorise the event in two divisions. The transonic accretion curve passing through the critical point situated away from the accretor, belongs to one class, and the transonic accretion curve passing through the critical point near the accretor belongs to the other one.
In the above figures, we choose different $\epsilon \mathscr{V}$ for different set of parameter values (The way of setting $\epsilon\mathscr{V}$ is described earlier). Now we try to figure out the dependence of $l$ on parameter values. Now we look at how the length $lc$ depends on the parameter values by keeping $\epsilon\mathscr{V}$ fixed. From equation \ref{lc}, 
\begin{equation}
lc=\frac{\epsilon\mathscr{V}}{|q|},
\end{equation}
where $q$ is the slope at the critical point (as discussed before). As $\epsilon\mathscr{V}$ is chosen to be constant, $l\propto\frac{1}{|q|}$. We find the variation in $\frac{1}{|q|}$ with the parameter values. 
\begin{figure}[hbtp]
\centering
%\captionsetup{margin=0.8cm}
\includegraphics[scale=0.38]{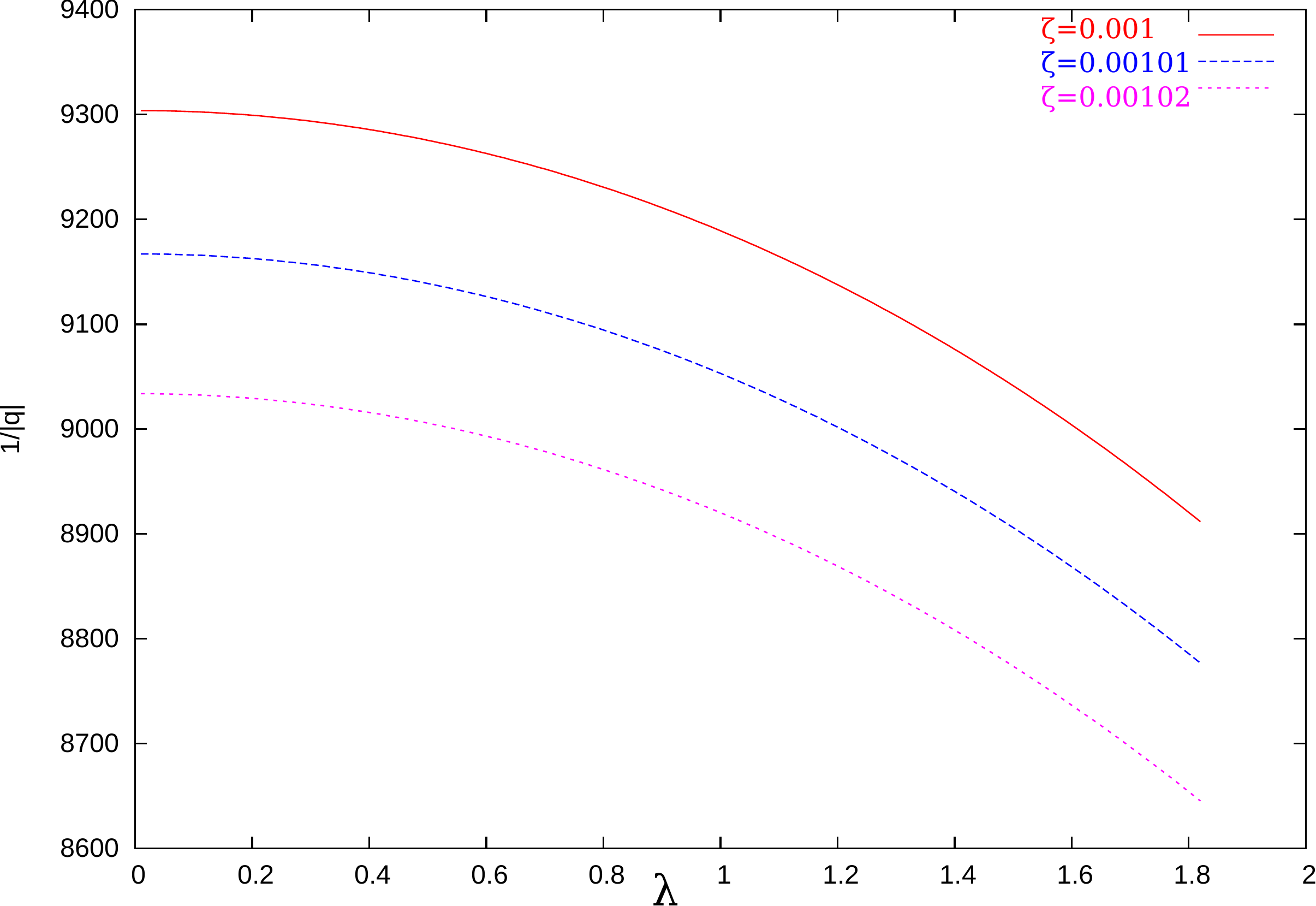} \\\includegraphics[scale=0.38]{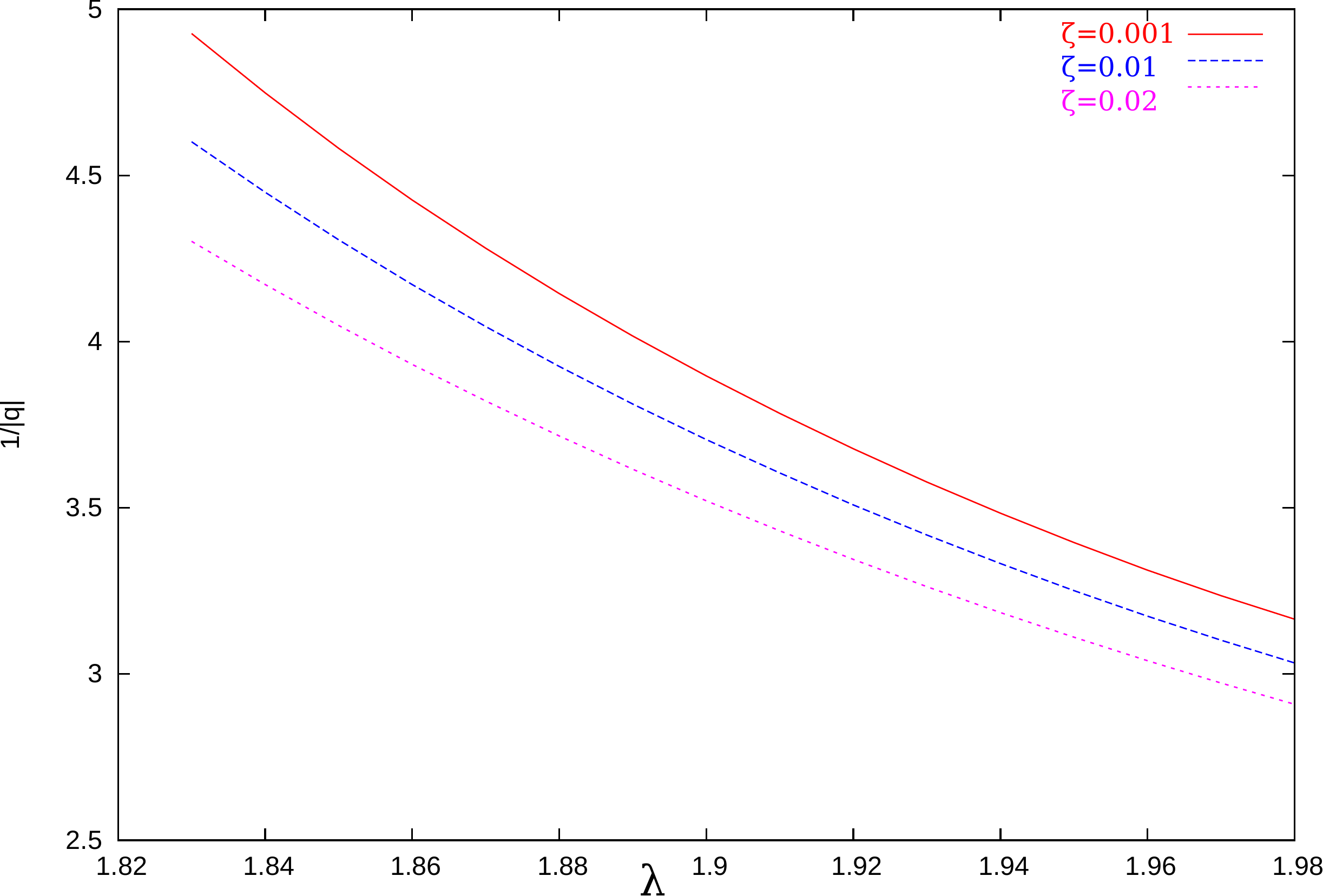}   
%\label{Figure1}
\caption{$\frac{1}{|q|}$ vs $\lambda$, region: accretion curve passing through the critical point situated away from the accretor (top) and  $\frac{1}{|q|}$ vs $\lambda$, region: accretion curve passing through the critical point situated near the accretor (bottom).}
\end{figure}
The figures clearly show that the length $l$ decreases with the increase of specific angular momentum and it increases with the decrease in the Bernoulli's constant.
\subsection{Variation with the Spin Parameter of the Pseudo-Newtonian Potential}
As the velocity profile \cite{z} (considering Pseudo-Kerr potential) looks qualitatively similar to the Pseudo-Schwarzschild case, we do not show the the variation of $log(l/lc)$ with $r$, rather we look at how black hole spin influences $lc$ in the following figures.
\begin{figure}[hbtp]
\centering
%\captionsetup{margin=0.8cm}
\includegraphics[scale=0.38]{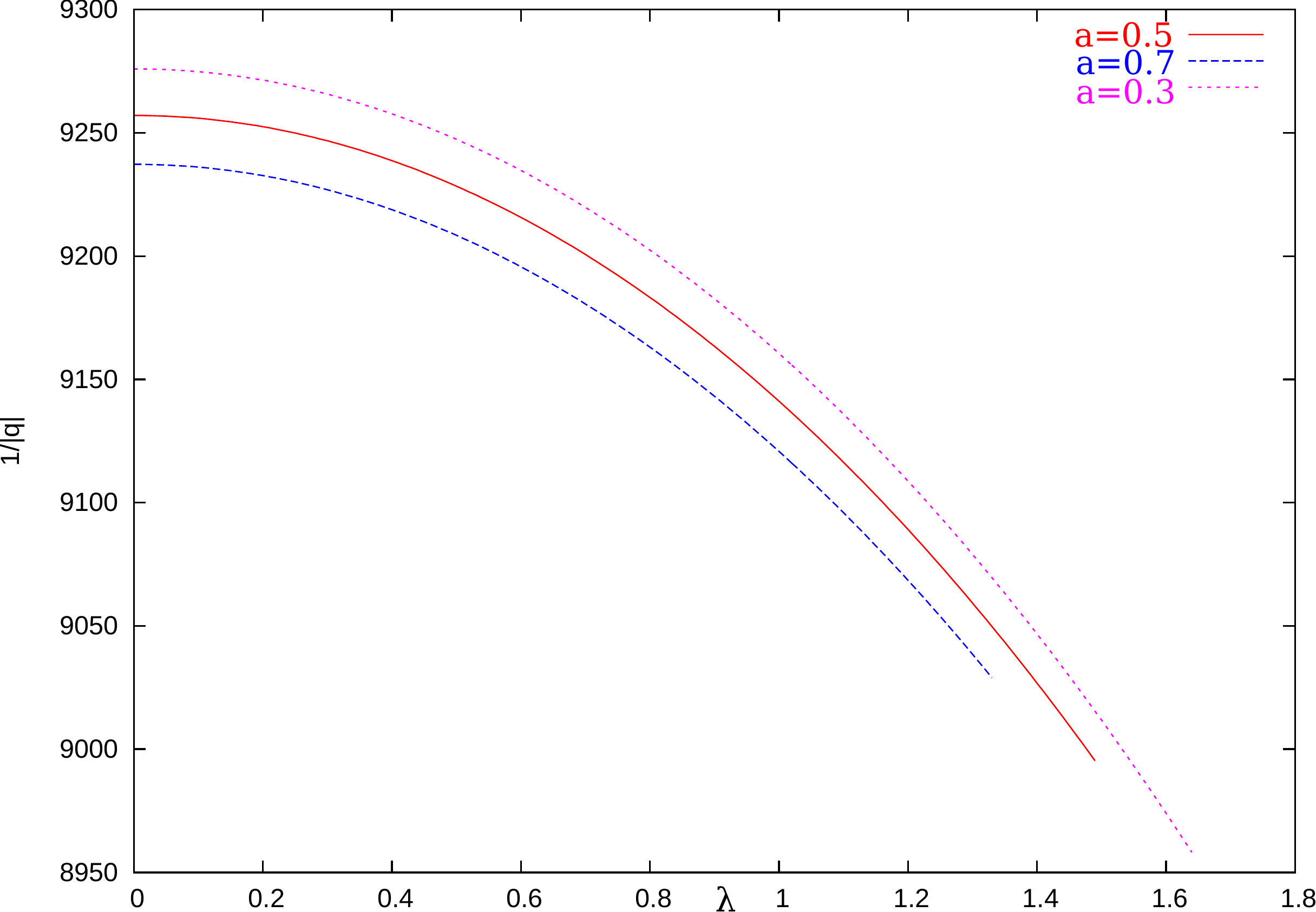} \\\includegraphics[scale=0.38]{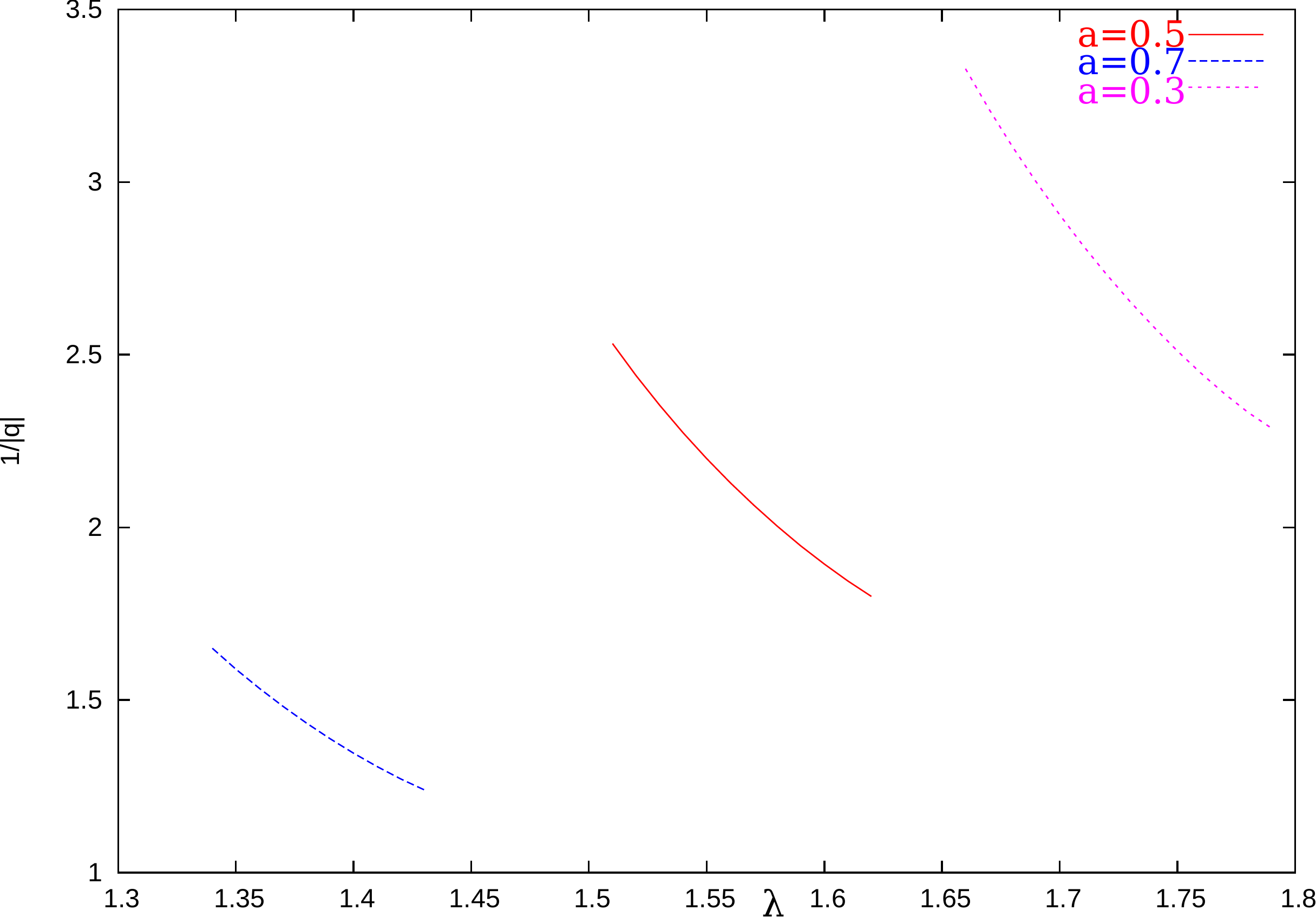}   
%\label{Figure1}
\caption{$\frac{1}{|q|}$ vs $\lambda$, region: accretion curve passing through the critical point situated away from the accretor; the curves corresponding to different values of $a$ stop at different $\lambda$ because that's the end of this kind of region in the parameter space (top) and  $\frac{1}{|q|}$ vs $\lambda$, region: accretion curve passing through the critical point situated near the accretor; the curves corresponding to different values of $a$ start at different $\lambda$ because that's the beginning of this kind of region (which is again the end of the mentioned region in the top figure) in the parameter space (bottom).}
\end{figure}
Variation of $lc$ with $\zeta$ is similar to as before. The figures indicate that $lc$ increases with the decreasing value of spin parameter, $a$. 
\section{Conclusions and Discussions}
 If the observer moving with the background flow perturbes the medium at his locations at different instants of time, and if the wavelength of such perturbations are small enough, within very short time within very short distance from that observer the wave propagates with same speed (the speed of sound in the static medium) along the possible directions. By `possible direction', we mean radially outward (from the accretor) and radially inward (towards the accretor) in the Astrophyical models of accreting blackl hole. Our aim has been to find the wavelength of such wave at different positions. We estimate a lengthscale ($=l$) such that wave having wavelength less than or equal to $l$, propagates with same speed along the available directions. Therefore, within a short time $\tau~(=\frac{l}{v_0}$, the time needed for that observer to cross that distance $l$. Within $l$, $v_0$ and $c_{s0}$ does not vary much), the observer can see the Eikonal wave propagating uniformly with speed $c_{s0}(r)$ within that distance, $l$. Our main aim in the work, has been to find the length $l$ to have an idea about the maximum wavelength of the Eikonal wave. Therefore, in the observer's reference frame, the Eikonal wave corresponding to this maximum wavelength, has time period, $T=\frac{l}{c_{s0}}$ (as evident from the dispersion relation \ref{dispersion R}). Hence, $\frac{T}{\tau}=M$ with $M$ being Mach number at that point. Hence, $\frac{1}{M}$ represents the number of full cycles, to be measured by the observer such that within the period of time $\tau$. Waves having wavelength less than or equal to $l$ has time period less than or equal to $T$. $\frac{1}{M}$ is the lower bound of number of cycles (=$\mathscr{N}$) to be observed by the moving observer to realize such effect. At the `dumbhole' horizon, $\mathscr{N}=1$. Thus even if the arbitrary choice of $\epsilon\mathscr{V}$ (depending on the choice of precision of the observer on the sensitivity of the background quantities with distance) makes $l$ arbitrary, we have found a concrete number which represents the minimum number of cycles of such Eikonal wave to realise such effect (the event of observing uniformly propagating sound along the relevant directions, as viewed by the observer moving with the background flow). The variation of $M$ with radial distance is discussed in details in several works \cite{z}-\cite{d1}.
From this work, we can also have some general qualitative beforehand idea (before doing any numerics) about how to find the variation of $l$ with $r$ from a given velocity profile in general. For example, in the spherically symmetric accretion (Bondi accretion), the speed of the background medium monotonically decreases with $r$ along the transonic accretion curve \cite{m} with asymptotically decreasing slope. Therefore, $l$ will monotonically increase becoming more steeper with $r$. Again, if there is a extrema in a velocity profile, in the figure of $l-r$, around that maxima or minima, there will be sharp peaks. 
%\appendix
\appendix
\section{Geometrical Acosutics} \label{A} 
About geometrical acoustics in the context of analogue gravity, there is a good discussion in the work \cite{c}, \cite{h1}. Here, we talk about it in a little different fashion for the convenience and completeness of our work.
Sound is treated as ray instead of a wave in the geometrical acoustics. The physical difference between wave and ray is that wave has the tendency to spread over in space (the nature of diffraction) but ray only follows a particular path (a curve in general) in space. This is quite similar to the discussion between wave nature and particle nature in Quantum Mechanics. Rays are defined as lines such that tangent to them at any point in space gives the direction of propagation. We know from the laws of diffraction that ideally if the wavelength of a wave is assumed to be zero then there is no diffraction. In the zero wavelength limit, resolution of an image (in case of light) is infinite. This limit of very short wavelength is the geometrical limit where the ray nature is a very good approximation.\\
Massless particle follows lightlike geodesic ($ds^2=0$) \cite{v}. We have wave equation for sound \cite{b}, \cite{c} in a moving inviscid locally irrotational inviscid flow, satisfied by linear perturbation in velocity potential $\psi'$
\begin{equation} \label{wave}
\partial_{\mu}(f^{\mu\nu}\partial_{\nu})\psi'({\bf{x}},t)=0 
\end{equation} 
where $f^{\mu\nu}$ in Cartesian coordinate is given by
\begin{equation}
f^{\mu\nu}=\frac{\rho_{0}}{c_{s0}^{2}}\begin{bmatrix}
-1 & \vdots & -v_{0}^{j} \\
\cdots&\cdots&\cdots\cdots \\
-v_{0}^{j}&\vdots & c_{s0}^{2}\delta^{ij}-v_{0}^{i}v_{0}^{j}
\end{bmatrix}
\end{equation}
where $i, j$ run over 1, 2, 3 representing three spatial dimensions.\\
The equation \ref{wave} represents a wave equation of the linear perturbation of the velocity potential.  
Comparing equation \ref{wave} with massless scalar field equation, one gets the acoustic metric \cite{b}, \cite{c}
\begin{equation}
g_{\mu\nu}=\frac{\rho_{0}}{c_{s0}}\begin{bmatrix}
 -(c_{s0}^{2}-v_{0}^{2}) & \vdots & -v_{0}^{j} \\
\cdots&\cdots&\cdots\cdots \\
-v_{0}^{j}&\vdots &\delta_{ij}
\end{bmatrix}
\end{equation} 
Now, we are asking that whether sound, satisfying wave equation \ref{wave}, follow the acoustic analogue of lightlike geodesic ( $ds^2=0$ where $ds^2$ is the acoustic metric) or not. For $ds^2=0$ between two neighbouring points in the spacetime, sound (linear perturbation) has to propagate in a certain direction, here the geometrical acoustics (short wavelength-high frequency limit) becomes useful. Therefore, we use basic equation for determining the direction of rays, by writing 
\begin{equation}
\Psi'=ae^{i\phi},
\end{equation}    
where the amplitude $a$ is slowly varying function of spacetime coordinate, and the phase $\phi$ is called eikonal. This approximation is called Eikonal approximation \cite{x}. The intensity (energy flux) of a steady ray never diminishes with distance (because it does not get diffracted) as it travels in space, that is why $a$ is assumed to be slowly varying function of spacetime coordinate. The wave vector, having the direction perpendicular to the constant $\phi$ surface at fixed time $t$, is given by
\begin{equation}
{\bf{k}}=\nabla\phi({\bf{x}},t).
\end{equation}
Angular frequency $\omega$ is defined as
\begin{equation}
\omega=-\frac{\partial\phi}{\partial t}.
\end{equation}
We can write the above expressions in tensor notation as
\begin{equation}
k_\mu=\nabla_\mu \phi,
\end{equation} 
where $k_0=\frac{\partial \phi}{\partial t}=-\omega$ and $k_i\equiv {\bf{k}}=\nabla\phi({\bf{x}},t)$.
Therefore, over a short distance from a point in space at an instant of time, the wavefront (the surface over which the phase is constant) of a ray is a plane. The normal to that constant phase surface at that point represents the direction of the ray at that instant of time. Hence, the eikonal wave, having high frequency (or short wavelength), is insensitive to the variation $f^{\mu\nu}$ within short distance in short time around a point in spacetime. Using the above expression of linear perturbation in velocity potential in the wave equation \ref{wave}, we find by equating the real and imaginary part:
\begin{eqnarray} \label{disperson} 
& k_\mu k^\mu=0\\ 
& \partial_\mu k^\mu=0. \label{dispersion2}
\end{eqnarray}
where $k^\mu=h^{\mu\nu}k_\nu$, $h_{\mu\nu}$ are components of the acoustic metric without the conformal factor in front of them, because equation \ref{disperson} and equation \ref{dispersion2} does not depend on the overall conformal factor due to the insensitivity of the eikonal wave on the variation of density and overall conformal factor in front of the physical acoustic metric. Therefore, we write down $h_{\mu\nu}$ and $h^{\mu\nu}$ as
\begin{equation} \label{hmn}
h_{\mu\nu}=\begin{bmatrix}
 -(c_{s0}^{2}-v_{0}^{2}) & \vdots & -v_{0}^{j} \\
\cdots&\cdots&\cdots\cdots \\
-v_{0}^{j}&\vdots &\delta_{ij}
\end{bmatrix},
\end{equation}
\begin{equation}
h^{\mu\nu}=\frac{1}{c_{s0}^2}\begin{bmatrix}
-1 & \vdots & -v_{0}^{j} \\
\cdots&\cdots&\cdots\cdots \\
-v_{0}^{j}&\vdots & c_{s0}^{2}\delta^{ij}-v_{0}^{i}v_{0}^{j}
\end{bmatrix}.
\end{equation}
With the above expressions in mind, equation \ref{disperson} refers to the dispersion relation of sound wave in high frequency-short wavelength limit (Eikonal approximation). Therefore, we find after some manipulation:
\begin{equation}\label{dispersion R}
\omega=\pm c_{s0}k+{\bf{v}}_0.{\bf{k}}
\end{equation}
We work with the `$+$' sign in the above equation because in a static medium, $\omega=c_{s0}k$.
The above equation is the dispersion relation of the wave in eikonal limit. $\omega$ is linear in $k$ due to the Lorentzian nature of the wave equation \ref{wave}. The first term in the right hand side is due to the Doppler effect (because of moving background medium). This term vanishes if the speed of the background medium tends to zero, where the emergent spacetime is analogous to Minkowskii spacetime.\\
Now using equation of rays \cite{x},
\begin{equation}
\dot{x_i}=\frac{\partial \omega}{\partial k_i}
\end{equation}
$\dot{x_i}$, the ith component of the group velocity of the wave, does not depend on wavelength of the sound, i.e., the medium is dispersive in nature. From the above equation,
\begin{equation}\label{vadd}
\frac{d{\bf{x}}}{dt}=c_{s0}\hat{n}+{\bf{v}}_0,
\end{equation} 
where $\hat{n}$ is the unit vector along ${\bf{k}}$. Therefore, we have
\begin{eqnarray}
& -(c_{s0}^{2}-v_{0}^{2})dt^{2}-2dt{\bf{v}}_0.d{\bf{x}}+d{\bf{x}}^{2} =0\\
&\Rightarrow ds^2=0.
\end{eqnarray}
Hence in the geometrical acoustics limit, sound follows null geodesic. The line element from equation \ref{hmn}, given by
\begin{equation}
ds^2|_{{\rm geometric}}=-(c_{s0}^{2}-v_{0}^{2})dt^{2}-2{\bf{v}}_0dt.d{\bf{x}}+d{\bf{x}}^{2}.
\end{equation}
This is the emergent metric in geometric acoustics regime. The conformal factor does not appear in the metric because null geodesic is insensitive to the conformal factor.\\
The equation \ref{vadd} can also be derived by directly using Galilean transformation (for more details see \cite{c}). Therefore, in the geometric acoustics limit, one does not even need to consider the wave equation \ref{wave}, i.e, one can start with velocity addition rule by Galilean transformation. Along the downstream the speed of sound in the laboratory reference frame gets added by the speed of the medium and along the upstream, the speed of sound gets subtracted by the speed of the moving medium.

\end{document}